\documentclass[showpacs]{revtex4}
\usepackage{graphicx}

\begin{document}

\title{Random matrix models for chiral and diquark condensation}

\author{Beno\^\i t Vanderheyden} \affiliation{Department of Electrical
Engineering and Computer Science, \\ Montefiore B{\^a}t.\ B-28,
University of Li\`ege, \\ B-4000 Li\`ege (Sart-Tilman), Belgium}

\author{A. D. Jackson} \affiliation{ The Niels Bohr Institute,
Blegdamsvej 17, DK-2100 Copenhagen \O, Denmark}

\date{March 2, 2005}

\begin{abstract}

We consider random matrix models for the thermodynamic competition
between chiral symmetry breaking and diquark condensation in QCD at
finite temperature and finite baryon density. The models produce mean
field phase diagrams whose topology depends solely on the global
symmetries of the theory. We discuss the block structure of the
interactions that is imposed by chiral, spin, and color degrees of
freedom and comment on the treatment of density and temperature
effects. Extension of the coupling parameters to a larger class of
theories allows us to investigate the robustness of the phase topology
with respect to variations in the dynamics of the interactions.  We briefly
study the phase structure as a function of coupling parameters and the
number of colors.

\end{abstract}


\maketitle


\section{Introduction}

Some thirty years ago, it was observed that dense and cold quark
matter might exhibit Cooper pairing as a result of an attractive
quark-quark interaction in the color antitriplet
channel~\cite{early-color-sup}.  More recent models, based on
non-perturbative effective interactions or on diagrammatic
calculations of single-gluon exchange interactions, indicate that a
color superconducting phase might develop pairing gaps as large as
$\Delta \sim 100~$MeV for quark chemical potentials on the order of
$300~$MeV~\cite{recent-color-sup}. This interesting possibility has
direct consequences on the physics of dense stars and is certainly
important for the determination of the phase diagram of nuclear matter
under extreme (ultrarelativistic)
conditions~\cite{color-sup-reviews,color-sup-lectures}.

Different order parameters have been proposed in the literature and
studied as a function of, e.g., the quark masses, the number of
flavors and colors, the quark chemical potentials, and
temperature~\cite{color-sup-reviews,color-sup-lectures}. In the limit
of QCD with two flavors of light quarks (the 2SC limit), the order
parameter has the form
\begin{eqnarray}
\langle \psi_{f\alpha\sigma}(p) \psi_{g\beta\sigma^{'}}(-p)\rangle 
= \Phi(p^2)\, \varepsilon_{fg}\,\varepsilon_{\alpha\beta 3}\,
\varepsilon_{\sigma\sigma^{'}},
\label{2SC}
\end{eqnarray}
where $p$ is a four-momentum and where we have displayed the flavor
($f$,$g$), color ($\alpha$, $\beta$), and spin ($\sigma$,
$\sigma^{'}$) indices. The tensors $\varepsilon$ ensures that the condensate
is antisymmetric in flavor, color, and spin. Color is broken from
$SU(3)$ to $SU(2)$, but the flavor symmetry $SU(2)_L\times SU(2)_R$ and
the baryon symmetry $U(1)_B$ remain intact.  In the other limit of
three degenerate flavors of light quarks, the favored order parameter
exhibits a coupling between color and flavor rotations (called
color-flavor locking or CFL) and has the approximate
form~\cite{color-sup-reviews}
\begin{eqnarray}
\langle
\psi_{f\alpha\sigma}(p) \psi_{g\beta\sigma^{'}}(-p)\rangle =\Phi(p^2)\,\,
\varepsilon_{fgA}\, \varepsilon_{\alpha\beta
A}\,\varepsilon_{\sigma\sigma^{'}}.
\label{CFL}
\end{eqnarray}
Color, flavor, and baryon symmetries are now broken down to
$SU(3)_{\mathrm{color}+L+R}\times Z_2$. Since both $SU(3)_\mathrm{L}$
and $SU(3)_\mathrm{R}$ are locked to $SU(3)_\mathrm{color}$, the CFL
condensate also breaks chiral symmetry.

\begin{figure}
  \includegraphics[height=0.25\textheight]{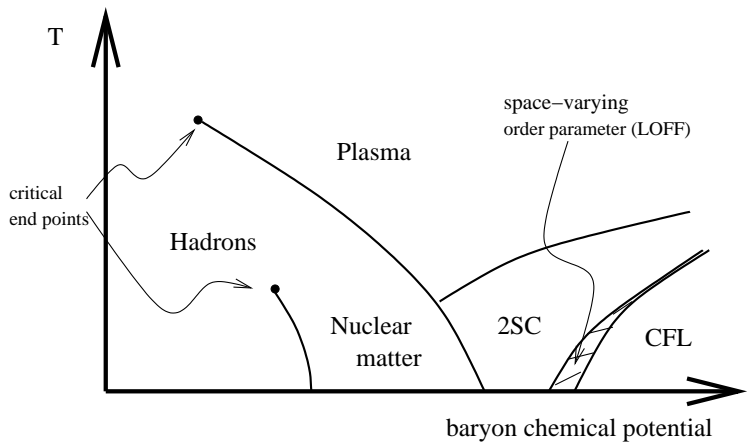}
  \caption{Sketch of the phase diagram for three flavor QCD with
  realistic quark masses. After Ref.~\cite{color-sup-lectures}.}
  \label{f:pd}
\end{figure}

Following Ref.~\cite{color-sup-lectures}, the conjectured phase
diagram for QCD with three flavors and realistic quark masses is given
in Fig.~\ref{f:pd}.  At asymptotically high densities, the scale is
set by the quark chemical potential, $\mu$. For $\mu \gg m_s$ ($m_s$
is the strange quark mass), the more symmetric CFL phase is
favored. As $\mu$ decreases, the increasing difference between $u$-
(or $d$-) and $s$-quark Fermi momenta weakens $\langle us \rangle$ and
$\langle ds \rangle$ condensates and eventually leads to a transition
to the 2SC phase~\cite{CFL-unlocking}, possibly via a so-called LOFF
phase characterized by a spatially varying gap~\cite{LOFF}.

Most of these results have not been confirmed by lattice simulations,
which are difficult to perform and interpret for QCD with three colors
at finite baryon densities.  The difficulty resides in the fact that the
fermion determinant in the partition function is complex for non-zero
$\mu$, and sampling weights are no longer positive definite. This
difficult problem represents a significant barrier to understanding the phase
structure of QCD in lattice simulations.

Random matrix theory offers models that are capable of distinguishing
those physical properties that are determined by global symmetries
from those that depend on the detailed dynamics of the
interactions. The random matrix Hamiltonian mimics the true
interactions by adopting a block structure that is solely determined
by the global symmetries under consideration. The matrix elements are
drawn on a random distribution, usually a Gaussian distribution, so
that the model can be solved exactly. Such an approach leads to
universal results that can be studied at two different levels. At the
microscopic level, the statistical properties of the lowest
eigenvalues of the Dirac operator are determined by the spontaneous
breaking of chiral symmetry, as indicated for example by the
universality of the spectral density near zero virtuality and related
sum rules~\cite{VerZah93,VerNPB94, VerPLB94,chRMT-reviews}. At the
macroscopic level, many of the properties of the phase diagram (such
as its topology and the presence of given critical lines or points)
are independent of the detailed form of the interactions and are thus
symmetry protected.  The random matrix approach treats low-lying
fermion excitations and, in its usual form, neglects their momentum
and kinetic energy. The resulting phase diagram is mean-field. Near
critical regions, it produces results similar to those obtained in a
Landau-Ginzburg approach~\cite{HalJacShr98,LG}.  In general, random
matrix models provide a useful tool for studying systems with
non-trivial phase diagrams.

This talk focuses on random matrix models that implement coexisting
chiral and color symmetries in QCD with two light degenerate flavors
(2SC). These models are extensions of the chiral random matrix models,  
which we introduce in Sec.~\ref{s:chiral}.  We present the
construction of the color and spin block structure of the interactions
in Sec.~\ref{s:expanded}. The phase diagram is discussed in
Sec.~\ref{s:pd}. This talk is a summary of previous work on the
question, see Refs.~\cite{VanJac00, VanJac01, VanJac03}.

\section{Chiral random matrix models}
\label{s:chiral}

We first consider chiral symmetry alone and turn to chiral random
matrix models~\cite{chRMT,VerPRL94}. For QCD in the sector of zero
topological charge with $N_f$ flavors and zero chemical potential, the
partition function is given as
\begin{eqnarray}
Z = \int\,{D} W \,\,\prod_{i = 1}^{N_f}\,\, {D}\psi_i^*\,{D}
\psi_i^{\phantom{*}} \, 
\exp\left[i \sum_{i= 1}^{N_f} \, \psi^*_i \, {\cal D} \, \psi_i
\right]\,\exp\Big(- \frac{n \beta \Sigma^2}{2}\,{\rm Tr}[W W^\dagger]\Big),
\label{Z}
\end{eqnarray}
where $W$ is an $n \times n$ matrix which models the interactions. Its
elements are drawn on a Gaussian distribution of mean zero and inverse
variance $\Sigma$.  Here, $n$ is a measure of the number of low-lying
degrees of freedom and is to be taken to infinity at the end of the
calculations (i.e., in the thermodynamic limit). In the chiral limit,
$m=0$, the Dirac operator, ${\cal D}$, has a block structure imposed
by the chiral symmetry of QCD, $\{D,\gamma_5\}=0$. In the basis of the
eigenstates of $\gamma_5$, this leads to the block structure
\begin{eqnarray}
{\cal D}= \left(
\begin{array}{cc}
0 & i W\\
i W^\dagger & 0 \\
\end{array}
\right).
\label{Dirac}
\end{eqnarray}
For random matrix models of QCD with $SU(3)$ and fermions in the
fundamental representation, $W$ is complex. This choice corresponds to
the chiral unitary ensemble, which is characterized by a Dyson index
$\beta = 2$. The QCD Dirac operator in $SU(2)$ with fermions in the
fundamental representation satisfies an additional anti-unitary
symmetry,
\begin{eqnarray}
[C(\sigma_2)_{\mathrm{color}} K,i {\cal D}] = 0,
\label{antiunitary}
\end{eqnarray}
with $\left(C(\sigma_2)_{\mathrm{color}} K \right)^2=1$.  ($C$ is the
charge conjugation operator, $\sigma_2$ is the antisymmetric color
matrix, and $K$ is the complex conjugation operator.)  This implies
that it is possible to find a particular basis of states in which
$i{\cal D}$ is real. Accordingly, $W$ is chosen real in chiral random
matrix models for $SU(2)$, and this leads to the chiral orthogonal
ensemble with an index $\beta = 1$~\cite{VerPRL94}. For fermions in
the adjoint representation of the gauge group and any number of
colors, the Dirac operator obeys the anti-unitary symmetry $C^{-1}K$ with
$(C^{-1}K)^2 = -1$. This leads to the symplectic chiral
ensemble with $\beta=4$ and quaternion real matrix
elements~\cite{chRMT-reviews, VerPRL94}. We will not consider this ensemble
further in this talk.

The essential difference among the ensembles lies in the number of
independent random variables that are allowed per matrix element.
(I.e., A complex number has two degrees of freedom; a real number has
only one.)  This difference is important in determining the
statistical properties of the Dirac operator and, in turn, affects the
phase diagram.

\section{Random matrix models with an expanded block structure}
\label{s:expanded}

In order to study the competition between chiral, $\langle \bar q q
\rangle$, and diquark, $\langle q^T q \rangle$, condensates, we introduce
an explicit dependence in the spin and color quantum numbers appearing
in the 2SC order parameter of Eq.~(\ref{2SC}). We are thus lead to a
Dirac operator with the chiral block structure of Eq.~(\ref{Dirac}),
where $W$ now has the following expanded color and spin sub-block
structure:
\begin{eqnarray}
W = \sum_{\mu = 0}^3 \sum_{a = 1}^{N_c^2 - 1} \lambda_a \otimes \sigma_\mu
\otimes A_{\mu a}.
\label{expanded}
\end{eqnarray}
Here, the deterministic matrices represent spin and color degrees of
freedom: $\sigma_\mu = (1, i \vec{\sigma})$ with $\vec{\sigma}$ the
Pauli matrices, whereas $\lambda_a$ are  color matrices (Gell-Mann
matrices for $N_c=3$). The random matrices, $A_{\mu a}$, are $N\times
N$ and represent gluon fields. They are chosen real. Their elements
are drawn on a Gaussian distribution with an inverse variance that is
independent of $\mu$ and $a$ in order to respect the Lorentz and
$SU(N_c)$ invariance in the vacuum.

Including a quark chemical potential, $\mu$, a quark mass, $m$, and
a temperature dependence, the partition function is now written as
\begin{eqnarray}
 && Z =  \int\,{D}\psi_1^\dagger\,{D}\psi_1^{\phantom{\dagger}} \,
 {D}\psi_2^*\,{D}\psi_2^T\,\,\{\prod_{\mu a} DA_{\mu a} \}
\,\exp\left(- 2 N \Sigma^2 \sum_{\mu a}{\rm Tr}[A_{\mu a}\,A^T_{\mu a}]\right)
\nonumber
\\
&\times &
\exp\left[i
\left(
\begin{array}{c}
\psi_1^\dagger \\
\psi_2^T \\
\end{array}
\right)^T
\left(
\begin{array}{cc}
i {\cal D} + i m + C_+ & 0 \\
0 & -i {\cal D}^T - i m - C_-^T\\
\end{array}
\right)
\left(
\begin{array}{c}
\psi_1 \\
\psi_2^*
\end{array}
\right)
\right],
\label{Zcolor}  
\end{eqnarray}
where the dependence on the external parameters is given by $C_\pm =
(\mp \sigma_3 \pi T + i \mu) \gamma_4$ with 
\begin{eqnarray}
\gamma_4 = \left(\begin{array}{cc} 0 & 1 \\ 1 & 0 \\ \end{array}\right).
\label{gamma_4}
\end{eqnarray}
Note that we have used a Gorkov representation and have transposed the
flavor-$2$ fields so that the diquark condensate appears as an
off-diagonal component of the inverse Dirac operator.

Several comments are in order regarding the $\mu$ and $T$
dependences. The $\mu$ dependence mimics the $\mu \psi^\dagger
\gamma_4 \psi$ term in the Euclidean QCD
Lagrangian~\cite{HalJacShr98}. The $T$ dependence is based on two
assumptions. First, temperature is introduced via a sum over the
fermion Matsubara frequencies, $\omega_n = i n \pi T$ with $n$ odd. As
random matrix theory only treats low-lying degrees of freedom, much of
the critical physics can be captured if one restricts the frequency
sum to its two lowest terms~\cite{JacVer96}, hence the two-dimensional
form ${\sigma_3 \pi T}={\mathrm{diag}}(\pi T,- \pi T)$. The second
assumption comes from the physical observation that the diquark order
parameter in Eq.~(\ref{2SC}) couples fields of opposite four-momenta.
We explicitly impose this coupling in Eq.~(\ref{Zcolor}) by taking
opposite Matsubara frequencies for each flavor. Different temperature
dependences have been proposed in closely related random matrix
models~\cite{KleTouVer03,KleTouVer04}; we discuss their relationship
to the present models in~\cite{VanJac05}.

\section{Phase diagrams}
\label{s:pd}

The partition function can be evaluated following a classical method.
An integration over $A_{\mu a}$ produces a four-fermion interaction
which can be Fierz-transformed to obtain the chiral and diquark
channel terms. The resulting four-fermion potential can be expressed
as a fermion field bilinear via a Hubbard-Stratonovitch
transformation. These successive steps yield~\cite{VanJac00}
\begin{eqnarray}
Z & \sim & \int d \sigma d \Delta \,\, e^{- 4 N \Omega(\sigma, \Delta)},
\label{Zfinal}
\end{eqnarray}
where the thermodynamical potential, $\Omega(\sigma, \Delta)$, is given
as
\begin{eqnarray}
\Omega(\sigma,\Delta) & = & A \Delta^2 + B \sigma^2 
- \frac{N_c - 2}{2}\, \sum_{\pm}\log\left( (\sigma + m \pm \mu)^2 + \pi^2
T^2 \right) 
- \sum_\pm  \log\left((\sigma + m \pm \mu)^2 + \pi^2 T^2 + \Delta^2
\right).
\label{Omega}
\end{eqnarray}
Here, $\Delta$ is the auxiliary field associated with the diquark
channel (with condensates $\langle \psi_{2R}^{T}
(i\sigma_2)_\mathrm{spin}\, \lambda_2 \psi_{1R} \rangle$ = $\langle
\psi_{2L}^{T} (i\sigma_2)_\mathrm{spin}\, \lambda_2 \psi_{1L}
\rangle$), while $\sigma$ is related to the chiral channel (with
condensates $\langle \psi_1^{\dagger} \psi_1^{\phantom{*}} \rangle = -
\langle \psi_2^{T} \psi_2^{*}\rangle$). The partition function can be
treated exactly in the thermodynamic limit $N\to \infty$ by a saddle
point method. The equilibrium values of the field then satisfy the two gap
equations
\begin{eqnarray}
\frac{\partial \Omega}{\partial \sigma} & = & 0, \\
\frac{\partial \Omega}{\partial \Delta} & = & 0,
\end{eqnarray}
which constitute a system of polynomial equations that can be solved
analytically or numerically.

The form of the potential in Eq.~(\ref{Omega}) has a straightforward
interpretation. The quadratic terms correspond to the energy cost for
having constant auxiliary fields. The logarithmic terms are to be
related to the single quasiparticle spectrum in a given set of fields
$\sigma$ and $\Delta$. Keeping in mind that the fermion four-momenta
are neglected, these energies are given as $\varepsilon = ((\sigma + m
\mp \mu)^2 + \Delta^2)^{1/2}$ for the two pairing colors (with the
$\mp$ sign standing for either quark or antiquark excitations). For
the $N_c - 2$ unpaired colors, $\varepsilon = |\sigma + m \mp \mu|$.

\begin{figure}
  \includegraphics[height=0.25\textheight]{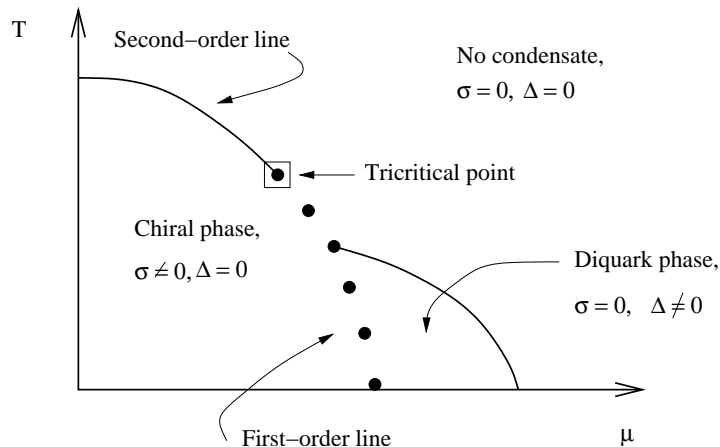}
  \caption{Phase diagram in the random matrix model for QCD with three
  colors and massless quarks.}
 \label{f:rm-pd-qcd}
\end{figure}

We now turn to a chiral symmetric theory with $m=0$.  Note that,
because it is possible to rescale all fields and external parameters
by a constant, the topology of the phase diagram depends only on the
ratio $B/A$. This is in fact a ratio of the Fierz constants in the
chiral and diquark channel, and it measures the relative importance of
the two symmetries. For example, small ratios are obtained with $A\gg
B$; the energy cost of $A\Delta^2$ then prohibits diquark
condensation. Similarly, large ratios $B/A$ disfavor chiral symmetry
breaking. For $N_c = 3$, the interactions in Eq.~(\ref{expanded}) lead
to a ratio $B/A = 3/4$, which corresponds to the phase diagram of
Fig.~\ref{f:rm-pd-qcd}. One observes a low-density chiral broken
symmetry phase separated from a higher density diquark phase by a
first-order line. As the temperature is raised, the diquark phase
makes a transition to the chiral symmetric phase across a second-order
line. The transition from broken to restored chiral symmetry is
second-order at low densities and first-order at intermediate
densities. The two lines are separated by a tricritical point, in the
vicinity of which the thermodynamical potential reduces to a $\phi^6$
theory~\cite{HalJacShr98,VanJac00}.

It is interesting to ask how the phase structure evolves if one
changes the channel couplings $A$ and $B$ from the values
representative of QCD. To this end, we have considered Hermitean Dirac
$i{\cal D}$ operators spanning an exhaustive set of combinations of
helicity, spin, and color block structures. Remarkably, this set
produces ratios $B/A$ in the {\em bounded} range $[0,N_c/2]$. This
result differs from what would have been obtained in a pure 
Landau-Ginsburg approach, where there is no {\em a priori} knowledge
of existing constraints among the coefficients of the effective
thermodynamical potential. Here, however, having started at a more
microscopic level, we are capable of discovering possible bounds on
the coupling ratios. Each ratio corresponds to a separate phase
structure. The major conclusions resulting from the study of phase
diagrams for ratios in the allowed range are as follows:
\begin{itemize}
\item 
there is only a finite number of different topologies. As $B/A$ is
varied continuously, the evolution from one topology to another is
marked by the emergence or the vanishing of new critical points or
lines;
\item 
it takes moderate --- but finite --- alterations of the theory to
depart from the topology of Fig.~\ref{f:rm-pd-qcd}. In that sense, the
phase structure of Fig.~\ref{f:rm-pd-qcd} is protected by symmetry.
\end{itemize}

\section{Discussion}
\label{s:discussion}

\subsection{Comparison with a microscopic theory}

In order to appreciate the approximations involved in the formulation
of the random matrix model, consider the gap equation in the limit
$\sigma = 0$,
\begin{eqnarray}
\frac{\partial \Omega}{\partial \Delta} = 0 \Rightarrow A\,\Delta = 
\frac{2 \Delta}{\Delta^2 + \mu^2 + \pi^2 T^2},
\label{Delta-gap-RM}
\end{eqnarray}
and compare with that obtained in a microscopic mean-field theory such
as that of Ref.~\cite{NJL}, based on an effective interaction
modeled by that induced by instantons. Approximately, we find
\begin{eqnarray}
\Delta \simeq  G \int \frac{d^4 p}{(2\pi)^4}\,
\left(
\frac{2 \Delta}{\Delta^2 + (p-\mu)^2 + p_4^2}+
\frac{2 \Delta}{\Delta^2 + (p+\mu)^2 + p_4^2}
\right) \ ,
\label{Delta-gap-NJL}
\end{eqnarray}
where we have dropped the form factors needed for the convergence of
the integral.  It is also implied that the integral over $p_4$ is a
sum over Matsubara frequencies, $p_4 = n \pi T$ with $n$ odd.

Because we have neglected the fermion four-momenta, the gap equation
in the random matrix approach does not contain an integral. This has
two consequences. First, the right term in Eq.~(\ref{Delta-gap-RM})
does not exhibit the logarithmic divergence observed in
Eq.~(\ref{Delta-gap-NJL}) for $p \simeq \mu$ and $\Delta = 0$. This
divergence arises for values of $p$ near the Fermi
momentum in the limit $\Delta \to 0$ and implies that a non-zero
$\Delta$ must develop at all $\mu$.  In contrast, the diquark phase in
Fig.~\ref{f:rm-pd-qcd} does not exist for asymptotically high values
of $\mu$. The random matrix interactions saturate in the diquark
channel for $\mu$ larger than $\sim\Sigma$. The second consequence
of the absence of a momentum integral is that the gap, $\Delta$,
evolves monotonically as a function of $\mu$ in contrast to
results from a microscopic approach, see Refs.~\cite{NJL,Bla05}.

These discrepancies should not be considered as weaknesses of the
random matrix approach. In the first case, $\Delta$ vanishes in a
region where microscopic theories tend to produce small gaps which
probably do not survive fluctuations beyond mean field.  What random
matrix cannot reproduce, however, is the $\Delta \sim e^{-1/g}$
behavior as a function of the QCD coupling constant, $g$. This
behavior is due to the magnetic gluon interactions~\cite{Son99}.  The
random matrix approach is also unable to reproduce the unbounded
increase of $\Delta$ at asymptotically high $\mu$, which is related to
the running of $g$~\cite{largemu}. Both these features are predicted
from diagrammatic theory and are related to particular dynamic
processes. The second observed difference, the variation of gaps at
moderate values of $\mu$, arises from the detailed dynamics of the
interactions and depends on the choice of the regulating form
factors. This behavior is not dictated by symmetry, and it is thus no
surprise that it is not revealed by the random matrix approach.

\subsection{QCD with two colors}

In the limit of $N_c = 2$, the gauge group is pseudoreal, and quark and
antiquark states transform similarly under global color
rotations. They can be combined into spinors which obey an extended
flavor symmetry $SU(2 N_f)$ for which $\langle \bar q q \rangle$
mesons and $\langle q q \rangle$ baryons belong to the same
multiplets. The random matrix model (with $N_f = 2$) reproduces this
extended symmetry in the vacuum and its breaking pattern as a function
of $\mu$, $T$, and $m$. In the vacuum and in the chiral limit, the
thermodynamic potential depends on the condensation fields through
the combination $\sigma^2 + \Delta^2$. The extended $SU(4)$ symmetry
is here apparent since a state with $(\sigma,\Delta) = (\Sigma,0)$ is
indistinguishable from its rotated version with $(\sigma,\Delta) =
(0,\Sigma)$. For $m > 0$ and low temperature, the phase $(\Sigma,0)$
undergoes a second order phase transition to a diquark phase at $\mu_c
\simeq m_\pi/2$ where $m_\pi\sim (m\Sigma)^{1/2}$ is the pion
mass~\cite{VanJac01}. Many results of the random matrix approach agree
with chiral perturbation theory in this case~\cite{chPT}.

\subsection{Two more questions}

The random matrix model is a theory of low-lying modes.  We argued
earlier that restricting the sum over Matsubara frequencies leads to a
model which nicely captures the critical physics. The neglect of
high-energy modes leads however to unphysical results, such as a
negative baryon density and a variation of the chiral field as a
function $\mu$, both in a theory with $N_c = 2$ and in the region $\mu
< m_\pi/2$.  We have shown that the inclusion of appropriate
high-energy terms (which should not describe the critical physics),
either in the form of correction terms or as a sum over all Matsubara
frequencies, fixes these anomalies while leaving the topology of the
phase diagram intact~\cite{VanJac01}.

Another question is whether the interactions in Eq.~(\ref{expanded})
preserves the statistical properties of the eigenvalues of the Dirac
operator that are expected from chiral symmetry alone. These
properties should follow the predictions of the chiral unitary
ensemble for QCD with three colors and those of the chiral orthogonal
ensemble for QCD with two colors. This question is related to the
number of random degrees of freedom allowed for the matrix
elements. We have shown that, even if the interaction matrices are
complex for both $N_c = 3$ and $N_c = 2$, the deterministic spin and
color dependences lead to the properties that are expected for the
spectrum of the Dirac operator~\cite{VanJac03}.

\section{Summary}

We have considered random matrix models of QCD with two flavors that
are capable of developing chiral and diquark condensation. We have
studied the symmetry breaking patterns as a function of temperature
and quark chemical potential. The phase diagrams can be established
exactly from an analytical evaluation of the partition function. The
resulting thermodynamical potential has a straightforward
interpretation in terms of elementary excitations. Upon arbitrary
variations of the coupling constants in the two condensation channels,
the phase structure evolves continuously but can only adopt a fixed
set of topologies.

This study shows that random matrix theory provides useful tools for
studying systems with non-trivial phase diagrams and for distinguishing 
those properties that are protected by symmetry.





\end{document}